# Skyrmion-Excited Spin Wave Fractal Network

Nan Tang,[1] W.L.N.C. Liyanage,[2] Sergio A. Montoya,[3,4] Sheena Patel,[3,5] Lizabeth J. Quigley,[1] Alexander J. Grutter,[6] Michael R. Fitzsimmons,[2,7] Sunil Sinha,[5] Julie A. Borchers,[6] Eric E. Fullerton,[3,8] Lisa DeBeer-Schmitt,[9] Dustin A. Gilbert[1,2*]

[1] Materials Science Department, University of Tennessee, Knoxville, TN
[2] Department of Physics and Astronomy, University of Tennessee, Knoxville, TN
[3] Center for Memory and Recording Research, University of California, San Diego, La Jolla, CA
[4] Naval Information Warfare Center Pacific, San Diego, CA
[5] Physics Department, University of California, San Diego, CA
[6] NIST Center for Neutron Research, National Institute of Standards and Technology, Gaithersburg, MD
[7] Neutron Scattering Division, Oak Ridge National Laboratory, Oak Ridge, TN
[8] Department of Electrical and Computer Engineering, University of California, San Diego, La Jolla, CA
[9] Neutron Scattering Division, Oak Ridge National Laboratory, Oak Ridge, TN

* Corresponding Author (dagilbert@utk.edu)

**Abstract**

Magnetic skyrmions exhibit unique, technologically relevant pseudo-particle behaviors which arise from their topological protection, including well-defined, three-dimensional dynamic modes that occur at microwave frequencies. During dynamic excitation, spin waves are ejected into the interstitial regions between skyrmions, creating the magnetic equivalent of a turbulent sea. However, since the spin waves in these systems have a well-defined length scale, and the skyrmions are on an ordered lattice, ordered structures from spin wave interference can precipitate from the chaos. This work uses small angle neutron scattering (SANS) to capture the dynamics in hybrid skyrmions and investigate the spin wave structure. Performing simultaneous ferromagnetic resonance and SANS, the diffraction pattern shows a large increase in low-angle scattering intensity which is present only in the resonance condition. This scattering pattern is best fit using a mass fractal model, which suggests the spin waves form a long-range fractal network. The fractal structure is constructed of fundamental units with a size that encodes the spin wave emissions and are constrained by the skyrmion lattice. These results offer critical insights into the nanoscale dynamics of skyrmions, identify a new dynamic spin wave fractal structure, and demonstrates SANS as a unique tool to probe high-speed dynamics.

**Introduction**

Magnetic skyrmions are a type of chiral soliton which are of interest for their fundamental properties that arise from a non-trivial, geometrically-defined topology.[1-5] This topology emerges as a result of the magnetic moments curling into a continuous co-planar loop, bounded on its perimeter and at its core by out-of-plane moments with opposite polarities. As a result of this structure, skyrmions possess a variety of dynamic excitations, such as breathing and gyration modes.[6-9] While there has been intense interest in these excitations for *e.g.* high frequency skyrmionic devices,[10] direct, *in-situ* investigations of skyrmion dynamics have remained challenging due to the combination of their exceedingly small size (10's to 100's of nanometers), high-frequencies ($10^9$ Hz), and magnetic-only contrast.

As the skyrmions undergo dynamic motion spin waves are ejected and propagate into the interstitial spaces between the skyrmions. Specifically, as the skyrmion undergoes dynamic distortion the constantly changing magnetization is compensated by the emission of spin waves. In most magnetic systems, the wavelength distribution of the spin waves follows a dispersion curve relationship of the form $f \propto k^2$. However, recent works[11-13] have predicted that the spin waves from skyrmion dynamics manifest with a well-defined periodicity which approximately matches the diameter of the skyrmion. These spin waves propagate outward from the skyrmion and can facilitate interactions between skyrmions[14] and potentially drive collective dynamics, but the spin waves can also interact with each other to form a turbulent magnetic landscape. This work reports that the dynamic magnetic structure generated by these spin waves is well represented by a 2D fractal network. Fractals play an important role in mathematics and science and can be generally described as complex geometric structures consisting of parts which repeat on different length scales, e.g. self-similarity. Fractal structures occur throughout nature, including coastlines[15] branching plants,[16] and polymer networks,[17, 18] for example. Unlike the 'deterministic' fractals described in mathematics, such as the Cantor and Mandelbrot sets or the Sierpinski triangle, which possess exact and infinite self-similarity, these 'random' or 'natural' fractals have statistical self-similarity.[16] Previous works have reported fractal structures in spin wave crystals.[19-21] An element of randomness is inherent in natural fractal systems, which in this case is the lack of chirality control in the skyrmions, resulting in 12 unique gyration modes and corresponding spin wave emissions.

Altogether, the current works reports on the dynamic structure of the skyrmion lattice (topological solitons), the spin waves which are ejected during their gyration, and the resulting dynamic structure (2D fractal network). A microwave-frequency excitation is applied to a thin film of hybrid skyrmions, driving them into resonant gyration and causing spin wave emissions. The magnetic structure is captured using small angle neutron scattering (SANS); SANS has been a critical tool in the investigation of skyrmions,[22-25] spin waves,[26-31] and fractals,[32, 33] but, as an elastic scattering tool, is not typically used to investigate dynamics. At resonance the SANS pattern shows significant changes, including notably the emergence of a new diffraction feature at very small angles, which is attributed to the spin waves. The SANS pattern is well fit with a mass fractal model, implying the spin waves form a long-range coherent structure. The spin wave fractal is constructed of fundamental units which match well to the skyrmion and skyrmion lattice structure. Results are supported by magnetic spectroscopy in the form of ferromagnetic resonance (FMR) measurement. These results, interpreted in the context of a micromagnetic model, provide enhanced insights into skyrmion dynamics and also expand the capabilities of the SANS technique.

**Results**

*Hybrid Skyrmion Thin films*

Thin films with a nominal structure of Si/Ta(50 Å)/[Fe(3.6 Å)/Gd(4.0 Å)]$_{120}$/Ta(50 Å) were grown by sputtering as described previously[34, 35] and in the Methods section. Skyrmions in this system have a hybrid structure which can be described as a Bloch-type skyrmion at the midpoint along the length of the skyrmion tube, *e.g.* its equator, with flux-closure domains – which emulate Néel skyrmions – at either surface.[34-36] The flux-closure domains form to contain the dipole fields from the film,[37] reducing the magnetostatic energy; the orientation of the Néel structure will follow the dipolar fields, which extend in opposite directions on the film's top and bottom surfaces, defining opposite chiralities to the surface spin structures. The chirality of the Bloch region can be either clockwise or counter-clockwise and occurs with equal probability.[34] This structure provides for excellent stability – including at ambient temperatures and zero field – and unique opportunities for sophisticated magnetic dynamics. The skyrmion state was prepared with either short-range or long-range ordering using the field sequence described in the Methods section.

*Microwave Spectroscopy of Skyrmion Dynamics*

For the FMR measurement the skyrmion state was prepared without tilting the sample, *e.g.* as domains with local hexagonal ordering, but no long-range orientation. FMR measurements were performed using a 'flip-chip' geometry with a grounded coplanar waveguide, generating an in-plane microwave excitation, with an out-of-plane static magnetic field. Figure 1 shows a 2D heat map of the transmission spectra $S_{1,2}$ measured as a function of frequency, $f$, and magnetic field, $H$. The heat map shows two series of measurements, each of which started from the skyrmion state at remanence: for $H<0$ the measurement proceeded from $H = 0$ to $\mu_0 H = -300$ mT, and for $H > 0$ the measurement proceeded from $H = 0$ to $\mu_0 H = +300$ mT; strong absorption, corresponding to resonant magnetic excitations, is indicated by the yellow contrast. The absorption map can be identified by seven regions indicated as (I-VII). In Region I the skyrmions are stable in their hexagonally close packed lattice. The data show strong resonant absorption following a ray which starts at ($H = 0$, $f = 1.7$ GHz), and ends at ($\mu_0 H = 190$ mT, $f = 2.4$ GHz). In the above-described measurement configuration, these dynamics are expected to correspond to in-plane gyration modes of the skyrmions. Increasing the magnetic field beyond 190 mT into Region II, the frequency corresponding to resonance rapidly decreases with increasing field. Further increasing the field beyond 220 mT (Region III), the resonant frequency again increases in response to an increasing static field. In this region the sample is saturated, and resonance corresponds to coherent, long-range spin gyrations, *e.g.*, Kittel resonance. This implies that Region II is a transition region, in-which skyrmions are being annihilated.

In the negative field range -110 mT $< \mu_0 H < 0$ mT, identified as Region IV, no apparent resonant signal is observed. In this region, the skyrmions are expected to combine, forming a labyrinth-like worm domain structure.[34, 35] The disorder and weak symmetry breaking between the up- and down-oriented domains may result in the suppression of long-range spin dynamics. Decreasing the field to Region V (-110 mT to -190 mT), the dynamics reappear, albeit slightly weaker and broader than the positive field analogue (i.e. Region I). This suggests the skyrmions re-emerge in this field range, which is much lower than reported previously with SANS.[23] However, the SANS results identify the ordered skyrmion lattice, not the emergence of dilute, disordered skyrmions, which is consistent with previous X-ray microscopy results.[34, 35] Regions VI and VII are largely symmetric with their positive field counterparts (Regions II and III, respectively) and indicate a transition region and saturated state, respectively. These FMR results are consistent with the previous static SANS measurements[23] and provide a map of skyrmion resonant modes which is used to guide the field and frequency conditions for the subsequent SANS measurements. Conceptually, analogous out-of-plane measurements could be performed to excite the breathing mode but were not performed here since the high frequency would induce significant eddy currents and sample heating in the metallic films.

*Neutron Scattering from the Dynamic Skyrmion System*

The SANS measurements discussed in the main text were performed on the GPSANS instrument at the High-Flux Isotope Reactor and the results discussed in the Supplementary Material were taken on the vSANS instrument at the NIST Center for Neutron Research. For the measurements on GPSANS, five films were stacked for the measurement to increase the signal. These samples were prepared to be in a long-range ordered skyrmion lattice following a magnetic field sequence reported in the Methods.[23]

The sample was placed in a two-turn solenoid, with the film in-plane direction aligned with the solenoid axis, and the neutron beam and static field aligned with the film normal direction, shown in Supplemental Material, Fig. S1. The resonant frequency of the solenoid was measured to be 2.23 GHz using a network analyzer; measurements were performed driving the solenoid at resonance with static fields from 100 mT to 190 mT, indicated with a red line in Fig. 1. Using a 5 W amplifier, the field from the two-turn solenoid is calculated to be ≈0.03 mT. Four representative SANS patterns are shown in Fig. 2a-d, captured at sequentially higher magnetic fields. The SANS plots are discussed in terms of their structure, which consists of the momentum transfer vector – the radial component – and the azimuthal scattering angle – the angle of scattering in the plane of the detector. At $\mu_0H$=130 mT, the six-fold diffraction pattern is readily apparent in the SANS data, confirming the skyrmions are stable in their hexagonally ordered lattice with a long-range orientation and that the applied dynamic fields are not distorting the lattice in any detectable way. An asymmetry in the peak intensity is apparent in the images, with the intensity being suppressed in the lower right quadrant, which is a consequence of a slight misalignment of the sample tilt and applied field relative to the neutron beam. The background appears as low-intensity diffuse scattering typical of SANS measurements. Increasing the applied field to 150 mT, the six-fold pattern is still easily identified, but the background scattering, especially in the low-$q$ region, starts to increase. Further increasing the applied field to 180 mT, the six-fold pattern is hard to distinguish within the increasing, angle-independent background intensity. As the applied field finally increases to 190 mT, the six-fold pattern reemerges and the intensity of the low-$q$ feature decreases significantly. Notably, the hexagonal SANS pattern has the same orientation after passing through resonance, suggesting long-range order was preserved throughout the measurement sequence. The differences between these different states are highlighted in Fig. 2e, which shows the intensity versus azimuthal angle, integrated over a range $q$=0.0029 Å$^{-1}$ ± 0.0005 Å$^{-1}$; this range captures the scattering intensity of the hexagonal diffraction pattern and the underlying, angle-independent scattering. Interestingly, these data show that the resonance increases the total scattering intensity (*i.e.*, integrated intensity in six-fold skyrmion peaks plus the low-$q$ feature). The additional scattering intensity is presumably located at $q$≈0 in the static measurements and shifts to higher $q$ outside the beamstop in the dynamic case. However, this interpretation implies the presence of large real-space structures, which are not expected from the skyrmion gyrations. Identifying the 180 mT measurement as corresponding to magnetic resonance, the FMR spectrum in Fig. 1 is suggested to have a small difference in the field calibrations. This series of measurements indicate that the SANS signal is indeed sensitive to the resonant skyrmion dynamics, with clear distinctions between the on-resonance and off-resonance signals.

*Quantifying the Effects of Gyration on the SANS pattern*

Integrating the SANS pattern across all azimuthal angles at fixed radii ($q$) generates a plot of intensity (*I*) versus $q$, shown in Fig. 3(a). The *I* vs. $q$ data measured away from resonance ($\mu_0H$ = 130 mT) shows a well-defined peak at $q$≈0.003 Å$^{-1}$ corresponding to the hexagonal skyrmion diffraction pattern, but little else. This peak appears in all of the measurements and is modeled by a general Gaussian function, $A_1 e^{-\frac{(q-\mu)^2}{\sigma}}$, where $A_1$ is the Gaussian amplitude, $\sigma$ is the peak width, and $\mu$ is the central peak position. As the applied field is increased towards 180 mT, a broad scattering feature appears at low-$q$, which rapidly decays. This decaying feature cannot be reasonably fit by many of the expected functions, including a power law (representing standard Porod-like scattering) or second-order exponential ($e^{-\frac{q^2}{c}}$), which would represent quasielastic scattering. A Lorentzian-squared function of the form $(\frac{A_2}{(1/\xi)^2 + q^2})^2$, which has been previously used to model frozen spin textures and spin waves,[26-31] was also tested and again provided a poor fit, shown in Supplemental Fig. S4. A comparison of all the alternative models is shown in Fig. S5, and their goodness of fit in Fig. S6.

The best fitting model, shown in Figure 3, was a mass fractal model.[18] This model describes systems which have an extended, chain-like or networked structure, constructed out of approximately monodisperse fundamental units. Such structures are commonplace in nanoparticle systems, where the fundamental building blocks (the nanoparticles) aggregate by e.g. van der Waals forces to form extended networks. This model is also commonly used to describe polymeric chains in-which regularly structured monomers form extended, cross-linked networks. Reflecting this physical interpretation, the mass fractal model has four variables: radius of the building block ($R$), fractal dimension ($D$), cut-off length ($\gamma$) and scale factor ($A_3$). Applying a mass fractal+Gaussian model to the entire dataset, Fig. 3(a) shows a good fit, with the converged parameters shown in Fig. 3(b). By subtracting the 130 mT data as an off-resonance background, the mass fractal model is able to achieve a high-quality fit across the entire data range ($0.0007$ Å$^{-1}$ < q < 0.01 Å$^{-1}$), shown in Fig. 3(c). The parameters for the fits show that the mass fractal behavior is strongly active during resonance, as indicated by the scale parameter, $A_3$. The converged model shows that the radius of the fundamental building blocks ($R$) that make up the fractal structure is approximately 150 Å - 200 Å, corresponding to a diameter of 300 Å - 400 Å; this length scale is consistent with the magnetic correlation length identified in the Lorentzian model in the Fig. S4. This length is speculated to correspond to the size of the pockets resulting from constructive and destructive interference within the spin wave network and correspond to approximately half of the wavelength of the spin waves. Next, the cut-off length ($\gamma$), which indicates the largest distance between points within the fractal, is shown to be ≈500 Å for most of the measurements, but increases to 1500 Å at resonance. The last parameter in the mass fractal model is the fractal dimension, which is around 2, consistent with a structure that is largely confined to the plane of the thin film.

The mass fractal model is conventionally used to describe a system constructed of monodisperse structures ordered into a fractal network. In the current system, the correlation between the mass fractal intensity, $A_3$, and the magnetic resonance suggests that spin waves take up this role. However, spin waves generally follow an energy-dependent dispersion curve, with a continuum of wavevectors (and corresponding wavelengths), making them inherently polydisperse; the fractal model implies that the spin waves have a well-defined length scale. It is also notable that this length scale is similar to the radius of the skyrmion, previously measured in similar films to be ≈350 Å.[34] This correlation between the length scale of the spin wave and the skyrmion radius was recently described by Satywali et al.[11] This work proposes that, during dynamic gyration, a magnon-skyrmion bound pair forms with the maximum of the magnon eigenfunction coinciding with the skyrmion radius, and rapidly decaying at both larger and smaller length scales. Considering the data in this context, we revisit the cut-off length ($\gamma$), which is ≈500 Å for the off-resonant conditions, increasing to ≈1500 Å on resonance. At 500 Å, the fractal features are smaller than the skyrmion and could be largely bound within it. On resonance, $\gamma$ is nearly the size of the skyrmion lattice (2200 Å), and matches very well with the lattice spacing minus the skyrmion diameter (1500 Å), e.g., the interstitial space between skyrmions. This implies that the fractal now fills the space between the skyrmions. Satywali et al.[11] measured two dynamic modes, one low frequency mode which is largely bound to the skyrmion, and one high frequency which results from dynamics between skyrmions. Our data may be probing these two conditions: in the off-resonant case the excitation drives weak dynamics, leaving the skyrmions largely non-interacting (that is, not interacting through magnons) and behaving as Satywali's low-frequency mode, while in the resonant state the strong dynamics increase the strength of the spin wave emissions, resulting in coupling and the high frequency mode. What was not expected is that these ejected spin waves would form an effective fractal structure of their own.

*Dynamic Skyrmion and Spin Wave Structure*

In the described experimental configuration, the skyrmion lattice exists in the plane of the Gd/Fe thin film with a static magnetic field applied along the film normal direction, parallel to the skyrmion tubes, and an in-plane excitation field. This geometry excites gyration of the skyrmion tube with the orbit in the plane of the film. Previous simulations performed on conventional skyrmion tubes reported that the structure of the skyrmion during its resonant motion can be significantly distorted, especially for higher-order dynamics.[8, 35, 38, 39] In the current system, even the lowest energy gyration modes will be complicated by the three-part structure of the hybrid skyrmion. As discussed in the Introduction, an individual hybrid skyrmion is comprised of a Bloch type winding at the equatorial band and two Néel skyrmions at the ends of the skyrmion tube; the Néel skyrmions have opposite chiralities following the dipolar fields emanating from the skyrmion core. For the static state shown in Fig. 4(a), the in-plane magnetic components along the +/-x and +/-y directions are balanced, while the core is centered within the chiral winding. Applying the excitation field along the $x$ direction, the structure of the dynamic skyrmion is well defined at each extremum of the driving cycle, with the domains parallel/antiparallel to the excitation field expanding/contracting and displacing the core,[11, 40] as shown illustratively in Fig. 4(b)*i* and *iii* for the positive and negative extremum, respectively. Connecting the center of the chiral winding from each section reveals that the core has a corkscrew-like shape along the length of the tube. After the extremum of the excitation field, the skyrmion core undergoes damped precession following the Landau–Lifshitz–Gilbert equation.[41] Without a transverse field (in ±y) or symmetry-breaking energy term (such as the Dzyaloshinskii–Moriya interaction (DMI)) each of the Néel and Bloch regions may gyrate clockwise (CW) or counterclockwise (CCW). Illustrations of three CW/CCW modes are shown in Fig. 4 (b)*ii*; within each of these three dynamic structures distortions of the core and the overlap of parallel domains in each layer will result in differences in the exchange energy. The resonant frequency of the precession is determined by the material properties (e.g. the perpendicular anisotropy $K_U$, exchange stiffness $A$, and saturation magnetization $M_S$) and applied magnetic field. In the gyration mode, the core is expected to follow an approximately circular orbit around the equilibrium center, with one precession per cycle of the excitation. The structure of the skyrmion in the static and dynamic cases were verified using the object oriented micro magnetic framework (OOMMF),[42] shown in Fig. 4(c) and (d), respectively. The static model emulates the structure in Fig. 4(a), while the dynamic model is presented in the extremum state, similar to Fig. 4(b)*i*. Notice that the regions parallel to the applied field (in +x for this example) in the dynamic models have grown while antiparallel regions have shrunk, similar to the illustrations in Fig. 4(b).

Each of the dynamic modes shown in Fig. 4(b)*ii* are doubly degenerate since the gyration direction (CW and CCW) can be exchanged, and double again since the direction of the chiral winding in the Bloch region can be switched (shown as CW in the image). This results in 12 unique gyration modes which will exist in the system, as illustrated in Supplemental Materials Fig. S2 These degeneracies will not change the energy but will affect the spin wave emissions. Each of these modes will eject a unique time- and depth-dependent collection of spin-waves into the film as their magnetization is continuously changing in time. This distribution causes the local spin wave environment to be disordered between unit-cells of the skyrmion lattice or from skyrmion to skyrmion, depending on the local gyration modes. However, these spin waves will have some common features, such as relative phase, since each skyrmion is driven by the same excitation field.

The ejected spin waves interfere to generate a diffraction pattern in the neutron data, however, it is the variation of these spin waves which causes this diffraction pattern to correlate with the mass fractal model. Specifically, considering a system of identical skyrmions with identical gyrations, the spin waves and resulting interreference pattern will be identical within each of the skyrmion lattice unit cells. The resulting spin wave interference pattern will have translational symmetry matching the skyrmion lattice, giving rise to a diffraction peak which is commensurate with the skyrmion lattice itself. Considering the spin wave emissions as a mechanism to conserve angular momentum during the gyration, switching the gyration direction inverts the polarity of the spin waves along the direction transverse to the excitation field, shown in Fig. 4(e). It is important to note that there are only three energetically unique modes, changing the Bloch orientation or all CW/CCW gyration directions does not change the energy, but clearly each will have different spin wave emissions. This is illustrated in Fig. 4e, which shows two Néel caps with the same chirality, but opposite gyration directions, and correspondingly different spin wave emissions. It is also notable that the spin waves have matching frequencies, and hence wavelengths, and are in-phase – as determined by the excitation field. Since the skyrmion lattice is randomly decorated with skyrmions with different gyration modes, the resulting interference pattern loses translational symmetry. The resulting spin wave interference pattern will be continuous across the film and constructed from fundamental units commensurate with the spin wave periodicity.

As a final consideration, the periodicity of the ejected spin waves is discussed. The neutron scattering suggests that the spin waves have a periodicity of ≈800 Å, corresponding to a wave number of 78 µm$^{-1}$. Using a standard $f \propto k^2$ dispersion relationship,[43, 44] the frequencies necessary to access this range are >10 GHz, while the experimental results (and simulations below) show resonance at ≈ 2 GHz. This is likely a consequence of the geometry of the system. Specifically, the spin waves propagating through an irregular (non-uniform) magnetic structure, confined to a film much thinner than the wavelength of the spin waves, resulting in deviations from traditional spin wave physics. Previous works performed on B20 skyrmion thin films have predicted resonant modes which are strongly dependent on the thickness of the film and can be <2 GHz.[38, 45] These works included a Dzyaloshinskii Moriya interaction term which is not included in this system. However, Ref. [38] suggests that, for the thin-film limit, which is the case here, that the magnetostatic interactions dominate the spectrum. In an alternative perspective, the dispersion relationship for a generic magnetic film was calculated specifically for thin-films with perpendicular static fields, called a forward volume spin wave (FVSW).[46] In all of these models, long wavelength spin waves are identified which can be propagated at lower frequencies than in their bulk analogue.

Micromagnetic simulations of a hexagonally ordered lattice consisting of 16 skyrmions during resonant dynamic gyration were performed as discussed in Methods. While the simulation in Fig. 4 illustrates the distortion of a single skyrmion and spin wave emissions during dynamic gyration, these simulations focus on the collective behavior and spin wave interference. The simulated magnetic field and resulting net magnetization in the $y$-direction, $m_y$, are shown in Fig. 5(a) and (b), respectively. Applying a Fourier transform to $m_y$ identifies the resonance frequencies as 1.6 GHz, 3.6 GHz, and 4.8 GHz. The resonance at 1.6 GHz agrees well with our experimental observations in Fig. 1 and Fig. S8 and S9, which showed that the resonance is between 1.3 GHz and 2.4 GHz, depending on the sample and applied field. Next, the excitation frequency was fixed at 1.6 GHz, driving the system at resonance. After eight cycles the system is presumed to be in a stable dynamic state and the magnetization in the $z$- and $y$-directions are exported for visualization, Figs. 5c and 5d; videos of these dynamic modes are shown in the Supplemental Material. Similar images taken in the off-resonance condition (frequency of 1 GHz) are shown in Figs. 5(e) and 5(f), and videos provided in the supplement, for comparison. For the $z$-contrasted images, red (blue) indicates magnetization in $+z$ ($-z$); in the $y$-contrasted images red (blue) indicates magnetization in $+y$ ($-y$).

In the $z$-contrasted images the skyrmions at resonance are not circular and have a distribution of sizes when compared to the off-resonance images, which show circular structures with a regular size. Both images show the skyrmions in an ordered hexagonal lattice, which is consistent with the SANS results. Z-contrasted video of the dynamics confirm that the skyrmions gyrate, but also that there are additional modes which propagate along the boundary of the textures. The irregular shape and edge modes may indicate that the driving frequency is slightly off resonance or may be the result of spin-wave mediated interactions between skyrmions. Video of the off-resonance condition shows minimal motion from the skyrmions. Notably, the spin waves are not visible in the $z$-contrasted video or images.

In the $y$-contrasted images captured during resonance the boundaries of the skyrmion lattice are clearly visible – the large red and blue crescents – but there is also a turbulent background resulting from the spin waves. The spin wave interference pattern lacks the well-defined structure which exists in deterministic fractals, again as a result of this being a 'natural' fractal, generated by a statistical kernel. Video of this state shows that the spin waves move very rapidly across the entire simulation space. Notably, the sizes of the magnetic regions are approximately regular – there are very few red, blue or white regions which are exceedingly small or large. The radius of the regions can be measured from the images and is found to be ≈220 ±30 Å, which is in reasonable agreement with the SANS results, which identified the radius of the 'building blocks' as 150 Å – 200 Å. Again, images taken in the off-resonance condition do not show any appreciable magnetic contrast beyond the static skyrmion lattice.

**Discussion**

The spin wave fractal network presents a novel dynamic structure which depends sensitively on the local skyrmion configuration. This structure necessarily incorporates elements of chaos to generate a quasi-ordered network, with multiscale self-similarity, but a lack of explicit translational symmetry. These qualities may present unique opportunities in novel magnon-based spintronic architectures.[43, 47-51] Stochastic computing, as a specific example, relies on a component of uncertainty or randomness in its operation. Alternatively, writing the chirality, such as in a nanostructured lattice,[52] may allow some control over the stochasticity of the network structure.

Previous works have also demonstrated that the periodic spatial modulation of the magnetization which results from the skyrmion lattice represents a natural two-dimensional magnonic crystal.[49, 53, 54] Accordingly, the collective ordering enforces a specific magnon spectrum[55] which, in this case, is driven by the peaked spin wave emissions[11-13] from the gyrating skyrmions. The resulting magnetic configuration, revealed by SANS includes the original skyrmion lattice as well as a magnetic fractal structure. This fractal structure, being both aperiodic and self-similar, can be well described as a (spatial) quasicrystal. However, this structure has a discrete translational symmetry *in time*. That is, the structure regularly repeats itself in time, locked to the driving frequency of the excitation field. Recent works have shown that driven spin waves in magnetic thin films can result in space-time crystals.[56-58] A similar space-time crystal (crystalline in time, quasicrystalline in space) may exist here, but would necessarily require resolving the nanoscale phase of the spin waves, which is beyond the scope of the current work.

These *in situ* SANS results provide several insights into the dynamics of hybrid magnetic skyrmions: (1) the skyrmion lattice is preserved and does not change significantly passing through resonance, suggestive of their robust topological protection, (2) the skyrmion dynamics eject spin waves with an associated length scale of ≈800 Å which are well represented with a mass fractal model, (3) fitting with the mass fractal model suggests that the spin waves at resonance extend outside of the individual skyrmions to form an extended structure analogous to a fractal network, (4) the observed mass fractal model supports the simulation and theory predictions presented by Satywali *et al*. These last three observations provide new insights into the complex magnetic landscape which is generated during skyrmion dynamics and suggests the emergence of a macroscopic magnetic superstructure.

While the correlation of the FMR and SANS results and subsequent high-quality fit offer compelling evidence that these results are correlated, considerations of alternative models are provided in the Supplemental Material.

This result presents a unique insight into magnetic dynamics and spin waves which may be applied beyond skyrmions to other magnetic systems. Specifically, the qualities of phonons and spin waves (magnons) are conventionally measured using inelastic neutron scattering techniques such as triple axis spectroscopy[59] (TAS) and time-of-flight[60] (ToF) spectroscopy. In these techniques, the energy loss and change in momentum of the scattered neutron is measured relative to the structural (elastic) scattering peaks.

However, the momentum transfer presented in these techniques is different from the momentum transfer vector discussed in SANS diffraction measurements which were performed here; this distinction is understood in scattering jargon as little-$q$ (momentum gain or loss relative to an elastic scattering event) versus big-Q (momentum transfer vector, change in the momentum direction but not magnitude). The momentum transfer in TAS and ToF does not explicitly encode the real space structure of the phonon or spin wave, only its direction, energy and total momentum. These other qualities can be derived from the dispersion curve, but a direct relationship, such as $d=2\pi/q$ where $d$ is the real space periodicity, which correlates the real-space structure to the diffraction peak location in elastic measurements, does not exist. The SANS measurements performed here, measured at unusually low-$q$ from the spin waves at the elastic position and does encode their real space structure, which is exceptionally rare.

**Conclusion**

We have successfully observed the gyration mode in magnetic skyrmions, as well as their spin wave emissions and their collective structure, using SANS as an *in-situ* probe. The gyration modes for the skyrmion lattice, including resonant and off-resonant states, can be excited by an in-plane gigahertz magnetic field and an out-of-plane static field. Throughout the measurements, including on- and off-resonance, the excitation field has a negligible effect on the structure of skyrmion lattice. However, during resonant gyration, a strong enhancement of background intensity in the form of a low-$q$ decay function, increasing the total scattering intensity by more than four-fold, is reported. This function can be fit with a Lorentzian squared function, corresponding to a collection of disordered magnetic clusters, but is better fit using a mass fractal model. The mass fractal model, which represents systems constructed of monodispersed elements coordinated into a fractal network, is attributed to the interference pattern generated by spin waves ejected during gyration. Notably, the lack of significant changes in the scattering from the skyrmion lattice and precipitation of an ordered spin wave network are surprising results. These measurements provide surprising new insights into an emergent long-range structure within the skyrmion lattice during dynamic gyration and suggests that SANS could become a relevant technique to investigate high-frequency magnetization dynamics.

**Methods**

*Sample Fabrication*

Thin-films with a nominal structure of Ta(50 Å)/[Fe(3.6 Å)/Gd(4.0 Å)]$_{120}$/Ta(50 Å) and Ta(50 Å)/[Fe(3.4 Å)/Gd(4.1 Å)]$_{120}$/Ta(50 Å) were grown on a 1 cm × 1 cm Si (001) substrate using direct current magneton sputtering. Film growth was performed at room temperature in a 3 mTorr Ar atmosphere. An ultra-high vacuum (UHV) background pressure of ≈1×10$^{-9}$ Torr was achieved before deposition to improve the quality of the films. The Ta layers were grown as a seed/adhesion layer and a capping layer against corrosion.

A skyrmion state with local ordering is prepared by the following field sequence: (1) +300 mT out-of-plane static field to saturate the sample, which (2) is then decreased to 0, precipitating a labyrinth domain structure, then (3) apply an out-of-plane field of +195 mT, which precipitates the magnetic skyrmions. Following this sequence the skyrmions form domains with short-range (several micron) hexagonal ordering but no long-range coherent orientation.

A skyrmion state with long-range ordering is prepared by the following field sequence: (1) A saturating magnetic field (+500 mT) is applied at 45° relative to the film normal direction, which (2) is then decreased to 0, resulting in magnetic stripe domains; the orientation of the stripes is determined by the in-plane component of the magnetic field. (3) Then a +195 mT field is applied along the film normal direction, breaking up the stripes to form hybrid skyrmions. The skyrmions form a hexagonally packed lattice, now with a long-range orientation defined by the initial stripe phase.

*Ferromagnetic Resonance*

The FMR measurements were performed using a grounded coplanar waveguide (CPW) connecting to a vector network analyzer (VNA) system. This VNA system can provide an alternating current through the CPWs with a range of frequency from 500 kHz to 32 GHz, generating a magnetic field which encircles the stripline. The Fe/Gd thin film is mounted with the film facing the CPW, with a Kapton film electrically insulating the film from the CPW structure; the generated magnetic field is thus in the plane of the film. Then the waveguide is inserted into a physical property measurement system (PPMS), which is used to control the temperature (1.8 K to 400 K) and static magnetic field (-9 T to 9 T); this static magnetic field is along the film normal direction.

At resonance, this magnetic field excites the magnetic skyrmions into resonant dynamic modes; the energy transfer from the stripline to sample is identifiable in the electronic transmission through the stripline, $S_{1,2}$. Measurements were performed at frequencies between 1 GHz and 3 GHz, with static out-of-plane magnetic fields between 0 and 300 mT. This measurement concludes in the saturated state. Next, the skyrmion state is reset by the same preparation process and $S_{1,2}$ was again measured at static fields between 0 and -300 mT.

*SANS Measurements*

Small-angle neutron scattering (SANS) experiments were performed on the GPSANS instrument at the High-Flux Isotope Reactor and the vSANS instrument at the NIST Center for Neutron Research. Measurements on GPSANS were taken at the maximal sample-detector distance (20 m), using long-wavelength neutrons ($\lambda$=16 Å, $\Delta\lambda/\lambda$ = 0.13). Data from GPSANS were reduced using the GRASP software package;[61] data from vSANS were reduced using the IgorPro Software.[62] A static magnetic field and the neutron beam were aligned with the film surface normal direction. A two-turn solenoid was arranged with one turn on either side of the sample, causing the magnetic component of the excitation field to be in the plane of the film. The configuration is shown in Supplemental Material Fig. 1. The solenoid was driven by a 5 W RF amplifier, coupled to an RF generator. Measurements were performed at room temperature both by fixing the frequency and sweeping the magnetic field, and fixing the field and sweeping the frequency.

Generally, SANS is not used to investigate dynamics since. This is because the measured scattering, as performed here, is an integral of all structures generated during the dynamic motion. A stroboscopic approach can be used on many beamlines now, however, due to the wavelength distribution of the neutron beam, this technique is generally limited to kilohertz frequency dynamics. Recent efforts have allowed SANS instruments to capture energy transfer.[63-65] Other neutron techniques, such as triple axis spectroscopy (TAS), can analyze the energy of the scattered neutrons but struggle to access the small momentum exchange vector ($q$) necessary to probe the skyrmion length scale.[39, 66-70]

*OOMMF simulation*

Three dimensional micromagnetic simulations for Fe/Gd multi-layers were performed for an individual skyrmion in simulation volume of (x=304 nm, y=280 nm, z=106 nm) using a rectangular mesh (x=6 nm, y=6 nm, z=2 nm). The material parameters in the simulated block include the exchange stiffness (A=0.7×10$^{-11}$ J/m), the saturation magnetization ($M_s$=3.5165×10$^5$ A/m) and the uniaxial anisotropy the z-direction ($K_u$=1.1×10$^5$ J/m$^3$). This exchange stiffness is chosen to be slightly smaller than the one in Ni (0.9×10$^{-11}$ J/m), while $M_s$ is directly measured by magnetometry and $K_u$ is calculated using the hard-axis saturation field.[71] These material parameters resulted in the stabilization of the hybrid skyrmion without an external magnetic field, similar to the experimental results; a significant decrease or increase of the value of these material parameters was found to cause the skyrmion to collapse. The stabilized single skyrmion with these parameters has a diameter of ≈150 nm. The static state was prepared by applying a magnetic fields ($B_x$=0, $B_y$=0, $B_z$=10 mT).

Dynamic simulations were performed by applying a static magnetic field of 10 mT is along the skyrmion tubes and an oscillatory field along the *x*-axis of the form $A_{field} Sin(B\, t^3)$, where $A_{field}$ is the maximal amplitude of the oscillatory field, and *B* is a constant of order $10^{24}$. Using this form, the frequency of the sinusoidal field increases from zero to 5 GHz over a simulation time of ≈25 ns. At resonance, the skyrmion undergoes gyration, resulting in a net magnetization in the *y*-direction, $m_y$. The spin wave emissions were simulated by driving the skyrmion lattice at a single frequency, ($B_x$= $A_{field}$ Sin(ω t) mT, $B_y$=0, $B_z$=10 mT), where ω is the resonant frequency determined by the initial simulations.

**Data availability**

The data that supports this work and the findings of this study are available from the corresponding author upon request.

**Acknowledgements**

Student support, travel, materials and the high-frequency equipment were supported by the U.S. Department of Energy, Office of Science, Office of Basic Research Early Career program under Award Number DE-SC0021344. Work at UC San Diego was supported by the National Science Foundation, Division of Materials Research Award #:2105400. We appreciate the assistance of the sample environment team at NIST, especially Alan Ye, the instrument team, notably John Barker, Jeff Krzywon and Cedric Gagnon, and Dan Neuman for insightful discussion. Access to vSANS was provided by the Center for High Resolution Neutron Scattering, a partnership between the National Institute of Standards and Technology and the National Science Foundation under Agreement No. DMR- 2010792 and DMR-1508249. Support for Lizabeth Quigley was provided by the Center for High Resolution Neutron Scattering, a partnership between the National Institute of Standards and Technology and the National Science Foundation under Agreement No. DMR- 2010792. Sergio A Montoya acknowledges the support from the U. S. Office of Naval Research, In-House Laboratory Independent Research. A portion of this research used resources at the High Flux Isotope Reactor, a DOE Office of Science User Facility operated by the Oak Ridge National Laboratory.

**Author Contributions**

Project and experiment design was performed by M.R.F., J.A.B., E.F., S.S., L.D.-S. and D.A.G. Sample fabrication was performed by S.M., and ferromagnetic resonance measurements by N.T., W.L.N.C.L. and S.M., SANS measurements were performed over three experiments by N.T., W.L.N.C.L., S.P., L.Q., A.J.G., J.A.B., L.D-S., and D.A.G. Analysis of the SANS data was performed by N.T., M.R.F., S.S., L.D.-S., A. J. G, J. A. B, and D.A.G. Micromagnetic simulations were performed by D.A.G. The first draft of the text was written by N.T. and D.A.G. All authors made contributions to and approved the final text.

**Corresponding Authors**

Dustin A. Gilbert: dagilbert@utk.edu

**Competing interests**

The authors declare no competing interests.

**Figures**

*Figure 1*

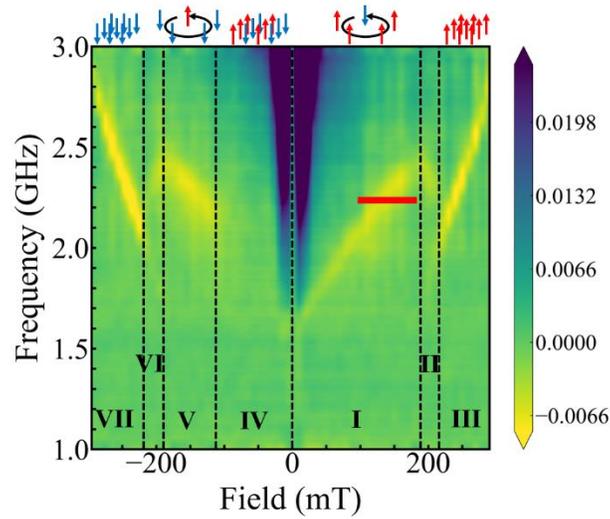

**Figure 1** Ferromagnetic resonance (FMR) of Fe/Gd multi-layers at room temperature, plotting the transmission coefficient $S_{1,2}$ as a function of frequency and out-of-plane magnetic field. Yellow indicates regions of high absorption. The labeled regions correspond to (I, V) locally ordered skyrmions, (II, VI) destruction of the skyrmion lattice, (III, VII) saturated state, and (IV) labyrinth domains. The red line indicates the nominal range of the SANS measurements. Contrast scale indicates the absorption parameter $S_{1,2}$, in dBm – a logarithmic scale.

*Figure 2*

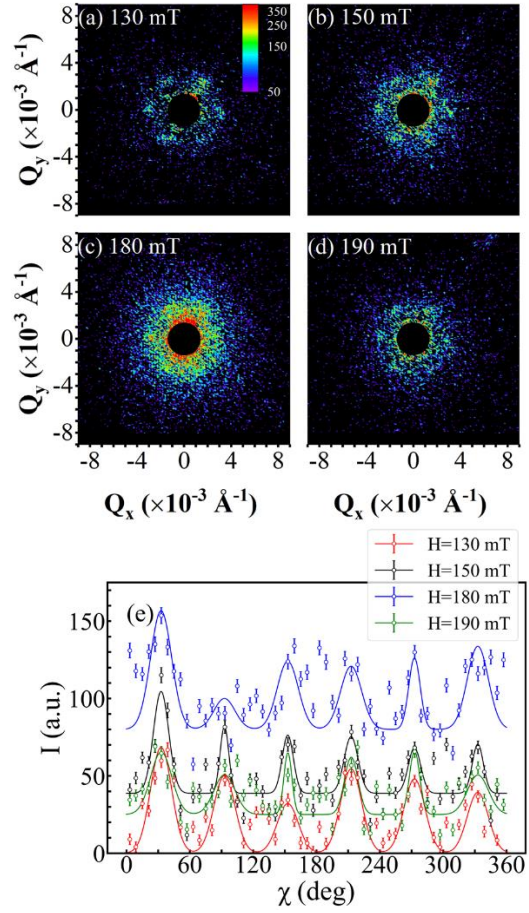

**Figure 2** 2D SANS pattern of magnetic skyrmions under different magnetic fields and excitation frequencies: (a) off-resonant state ($\mu_0H$ =130 mT, $f$ =2.23 GHz), (b) initial-resonance state ($\mu_0H$ =150 mT, $f$=2.23 GHz), (c) on-resonance state ($\mu_0H$ =180 mT, $f$ =2.23 GHz) and (d) post-resonance state ($\mu_0H$ =190 mT, $f$ =2.23 GHz). All 2D images are plotted with the same intensity scale. (e) Intensity vs azimuthal angle ($q$=0.0029 Å$^{-1}$, $\Delta q$=0.0005 Å$^{-1}$), 0° corresponds to the vertical position and the angle is measured in the clockwise direction.

*Figure 3*

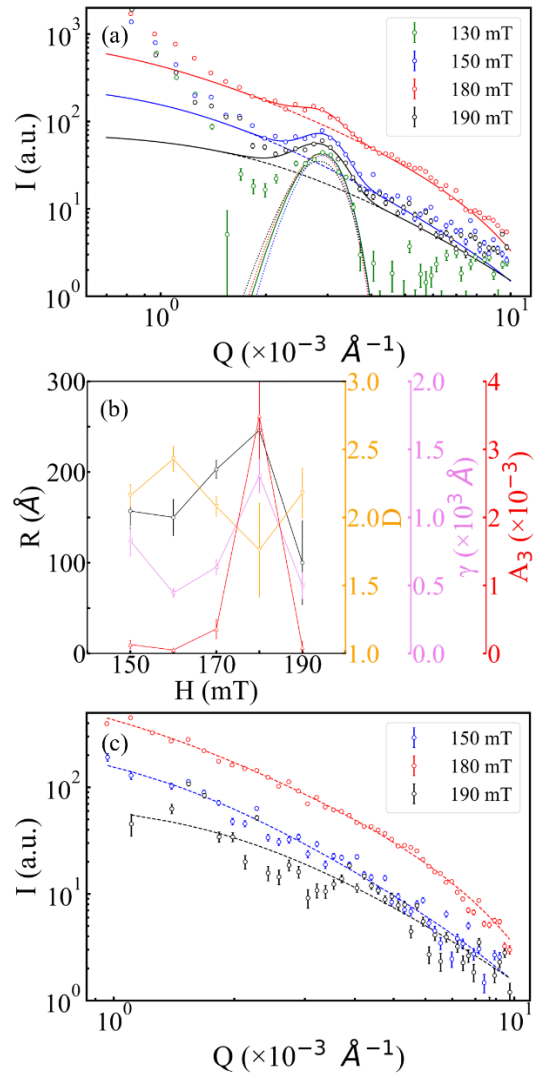

**Figure 3** (a) *I* vs *q* plots for the off-resonant (green), pre-resonance (blue), on-resonance (red) and post-resonance (black) states. The data are represented by points and the fitted models are shown as solid lines; the Gaussian component of the model is shown as a dotted line and the mass fractal model as a dashed line. (b) Converged fitting parameters for the mass fractal model: radius of the building block (*R*), mass fractal dimension (*D*), cut-off length (γ) and scale factor (*A3*). (c) I vs q plots for pre-resonance (blue), on-resonance (red) and post-resonance (black) states using the 130 mT data used as a dynamic-independent background, then fitted with the mass fractalmodel.

*Figure 4*

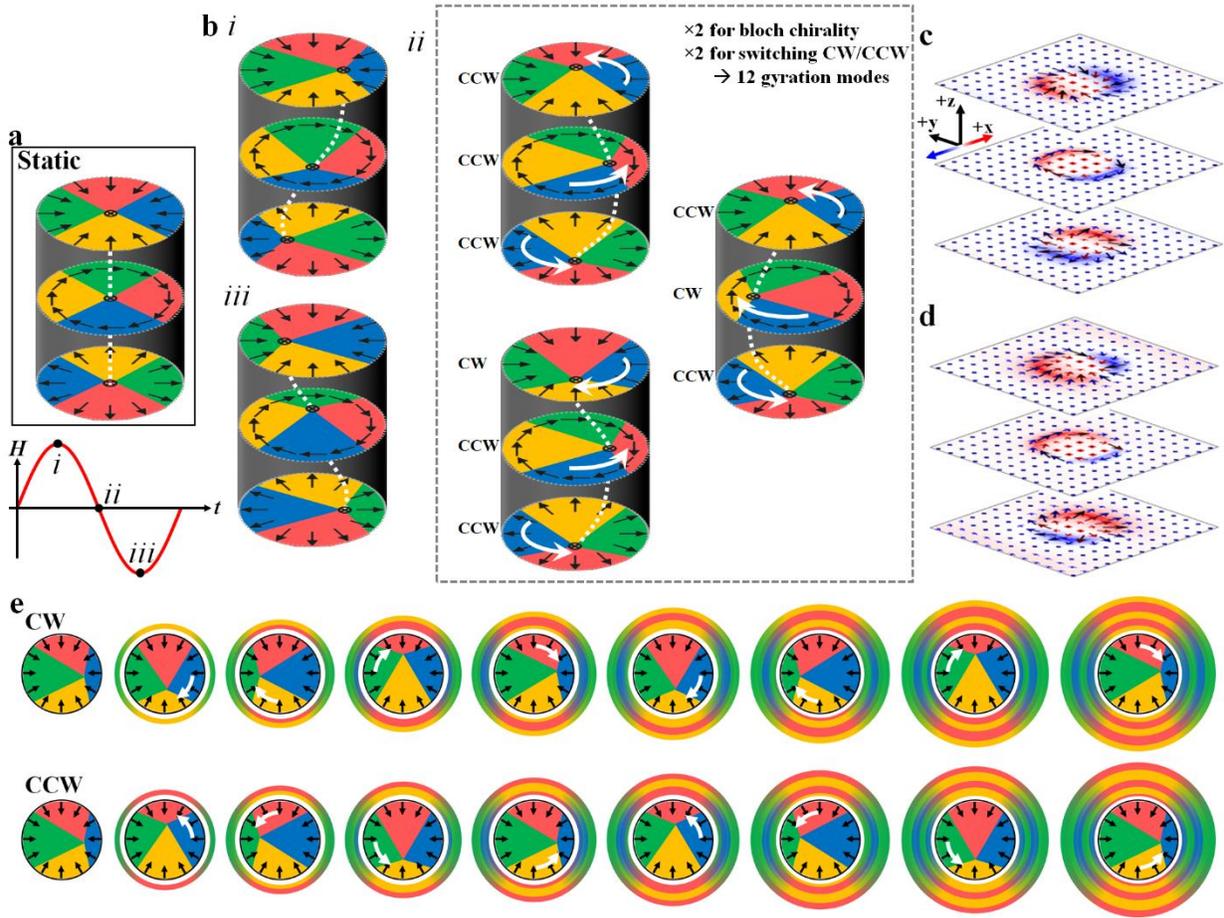

**Figure 4** Illustrative diagrams showing the (a) static and (b) dynamic gyration of a hybrid skyrmion. At the extremum of the excitation (panel (b), *i* and *iii*) the magnetic configuration is well defined, however, the gyration between these states can be either clockwise or counterclockwise, shown in panel (b)*ii*. Spins in the diagrams are indicated with black arrows, the approximate position of the core is traced in the dashed white line and the motion of the core illustrated with solid white arrows. Micromagnetic simulations of (c) the static and (d) dynamic state; the magnetic configurations are analogous to panels (a) and (b)*i*, respectively. Colors indicate the *x*-component of the magnetization, with red (blue) indicating magnetization in +x (-x). (e) Illustrations of the spin wave emissions from a Néel cap gyrating CW and CCW, where the color of the rings is determined by the changing areas of the skyrmion quadrants, and radiate outward from the skyrmion.

*Figure 5*

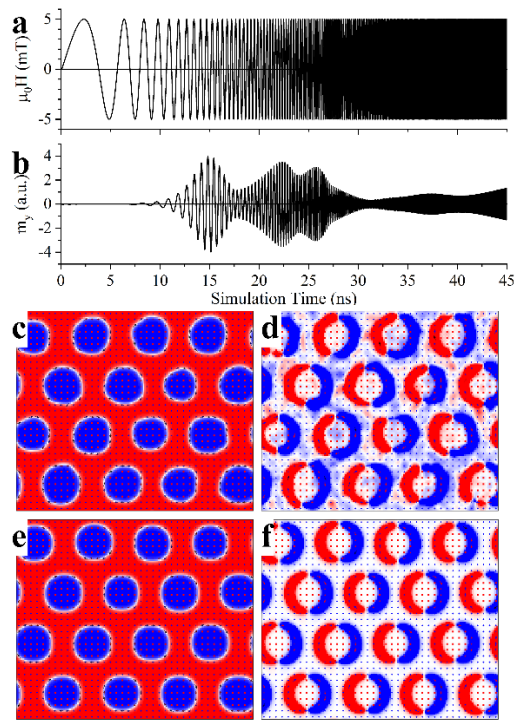

**Figure 5** (a) Applied in-plane excitation field, applied parallel to the x-axis, and (b) net magnetization along the y-axis. Resonance occurs at 1.6 GHz, 3.6 GHz and 4.8 GHz. Micromagnetic results showing the magnetization along (c) the out-of-plane z-direction and (d) in-plane y-direction, for an excitation frequency of 1.6 GHz; magnetization along the (e) z- and (f) y-axes, driven off-resonance at 1 GHz.